\documentstyle[12pt]{article}
\def\be{\begin{equation}}
\def\ee{\end{equation}}

\newtheorem{theorem}{Theorem}

\newtheorem{definition}[theorem]{Definition}

\begin{document}

\title{Localization in space for free particle \\ in ultrametric quantum mechanics}

\author{A.Yu.Khrennikov\footnote{International Center for Mathematical
Modelling in Physics and Cognitive Sciences, University of
V\"axj\"o, S-35195, Sweden, e--mail:
Andrei.Khrennikov@msi.vxu.se}, S.V.Kozyrev\footnote{Steklov
Mathematical Institute, Moscow, Russia, e--mail:
kozyrev@mi.ras.ru}}

\maketitle

\begin{abstract}
Free evolution for quantum particle in general ultrametric space
is considered. We find that if mean zero wave packet is localized
in some ball in the ultrametric space then its evolution remains
localized in the same ball.
\end{abstract}

\section{Introduction}

One of interpretations of ultrametric (in particular, $p$--adic)
quantum mechanics is the description of matter at Planck
distances. In the present paper we find that if mean zero wave
packet is localized in some ball in ultrametric space then its
evolution remains localized in the same ball. In this
interpretation quantum wave packet at Planck distances remains
localized in space even for free particle. This property has no
analogs in real quantum mechanics and characterizes ultrametric
quantum mechanics.

We consider ultrametric quantum mechanics with real time and
ultrametric space.  Quantum mechanics with ultrametric ($p$--adic
and adelic) space was considered in many works, see for example
\cite{VV213}--\cite{VVZ}. $p$--Adic and adelic models in string
theory and cosmology were developed \cite{VVZ}--\cite{bib16}. For
these kind of applications it is important that $p$--adic numbers
are non--archimedean and have no order, which corresponds to
presence of physically fixed Planck distance. Presence of a
natural distance scale is a typical property of ultrametric
spaces.

Physical consequences of violation of the Archimedes axiom (in
particular, for measurement theory) were discussed in
\cite{Andr1}, \cite{VVZ}. However, $p$--adic numbers still have
the group (and the field) structure. Existence of such structures
at Planck scales could be also questioned. In the present paper we
consider quantum model, for which space, in principle, does not
permit any group structure, and possesses the only property of
ultrametricity (which can be physically justified by existence of
fixed Planck distance).

\section{Free ultrametric particle}

We consider the ultrametric quantum particle propagating generic
ultrametric space, which in general has no any group structure.
The dynamics of the particle is described by the ultrametric
Schrodinger equation (\ref{uSe}) below. The main observation of
the present paper (which is valid also for the particular
$p$--adic case) is that ultrametricity of the space implies the
new phenomenon --- localized in space mean zero wave packets
remains localized for any moment of time. Here we assume that time
is real but the same observation is valid for the case when both
space and time are ultrametric.

The model under consideration is as follows (see the Appendix for
the details). Consider product $X\times R$ of ultrametric space
$X$ of the analytic type and the real line $R$, with the
coordinates $x$ and $t$ correspondingly. Consider ultrametric
pseudodifferential operator $D_x$ of the sup--type (see the
Appendix), acting on functions in $L^2(X,\nu)$, and the
ultrametric Schrodinger equation with real time \be\label{uSe}
\left(i\hbar\partial_t-\hbar^2D_x\right)\Psi(x,t)=0 \ee  Equation
(\ref{uSe}) describes free quantum particle in ultrametric space
and real time. One can also consider ultrametric quantum particle
in potential, described by the following ultrametric Schrodinger
equation
$$
\left(i\hbar\partial_t-\hbar^2D_x-U(x)\right)\Psi(x,t)=0
$$
In standard quantum mechanics to obtain localized in space
solution of the Schrodinger equation one have to consider
Schrodinger equation with potential. In ultrametric case we obtain
localized in space solutions already for the free case
(\ref{uSe}).

Existence of localized solutions of the ultrametric Schrodinger
equation is related to the fact that ultrametric
pseudodifferential operators (for example, the Vladimirov operator
of $p$--adic fractional differentiation) has bases of eigenvectors
with compact support. The example of bases of this kind are
ultrametric (in particular, $p$--adic) wavelets.

Let us first construct an example of localized solution of
(\ref{uSe}). Consider ball $I\subset X$ and the ultrametric
wavelet $\Psi_{Ij}(x)$ supported in this ball (see the Appendix
for the construction of ultrametric wavelets). Let this wavelet be
an eigenvector for $D_x$, corresponding to the eigenvalue
$\lambda_I$:
$$
D_x\Psi_{Ij}(x)=\lambda_I\Psi_{Ij}(x)
$$
Then the function
$$
\Psi(x,t)=e^{-i\hbar \lambda_I t
}\Psi_{Ij}(x)
$$
is the solution of the free ultrametric
Schrodinger equation with real time (\ref{uSe}). This solution is
supported on the ball $I$ in ultrametric space $X$. This solution
may be discussed as an ultrametric wave packet.

In general case, expand $\Psi(x)$ over ultrametric wavelets
\be\label{expand} \Psi=\sum_{Ij}\Psi_{Ij} \ee Since the function
$\Psi$ is mean zero, index $I$ in the expansion above (which
enumerates balls in $X$) runs over $I\subset B$.

Solution of the ultrametric Schrodinger equation (\ref{uSe}) with
initial condition (\ref{expand}) has the form \be\label{expand1}
\Psi(x,t)=\sum_{Ij,I\subset B}e^{-i\hbar \lambda_I t }\Psi_{Ij}(x)
\ee Since ultrametric wavelets $\Psi_{Ij}$ are supported in balls
$I$, function $\Psi(x,t)$ defined by (\ref{expand1}) is supported
in ball $B$. Integrating $\Psi(x,t)$ over $B$ we obtain zero. We
have the following theorem.

\begin{theorem}
{\sl Let $\Psi(x)$ be mean zero function in $L^2(X,\nu)$ with
compact support in some ball $B\subset X$. Then the solution
$\Psi(x,t)$ of the ultrametric Schrodinger equation (\ref{uSe})
with initial condition $\Psi(x)$ for arbitrary $t$ is mean zero
and has support in the same ball $B\subset X$. }
\end{theorem}

\noindent{\bf Remark}\qquad It is clear that the above theorem
will also be satisfied for the following ultrametric heat equation
$$
\left(\partial_t+D_x\right)\Psi(x,t)=0
$$
In the real case all localized at the initial time moment
solutions of the Schrodinger equation and the heat equation loose
localization. In the ultrametric case any localized in some ball
mean zero function conserves localization at any time moment.

\bigskip

In paper \cite{VV213} it was supposed that at the Planck distance
scale space and matter are described by $p$--adic quantum
mechanics. In this interpretation of ultrametric quantum mechanics
we get that at the Planck distance localized quantum wave packets
remain localized. This localization effect which will take place
even for free particle and has no analogs in real quantum
mechanics.

\bigskip

\noindent{\bf Remark}\qquad In the case, when both space and time
coordinates $x$ and $t$ are described by ultrametric spaces $X$
and $Y$ correspondingly, the quantum dynamics of the particle in
ultrametric space--time is described by the following ultrametric
Schrodinger equation \be\label{uSe1}
\left(AD_t-BD_x\right)\Psi(x,t)=0 \ee Here $D_x$ and $D_t$ are
ultrametric pseudodifferential operators of the sup--type, acting
on functions of $x$ and $t$.

Consider balls $I\subset X$, $J\subset Y$, and the ultrametric
wavelets $\Psi_{Ij}(x)$, $\Psi_{Jj'}(t)$, which are the
eigenvectors of $D_x$ and $D_t$ correspondingly:
$$
D_x\Psi_{Ij}(x)=\lambda_I\Psi_{Ij}(x)
$$
$$
D_t\Psi_{Jj'}(t)=\lambda_J\Psi_{Jj'}(t)
$$
Assume that $\lambda_I$, $\lambda_J$ are non zero. Then one has
$$
\left({1\over\lambda_J}D_t-{1\over\lambda_I}D_x\right)\Psi_{Ij}(x)\Psi_{Jj'}(t)=0
$$
This shows that the product of ultrametric wavelets is the
solution of the ultrametric Schrodinger equation (\ref{uSe1}) for
$A={1\over\lambda_J}$ and $B={1\over\lambda_I}$. The function
$\Psi(x,t)=\Psi_{Ij}(x)\Psi_{Jj'}(t)$ is compactly supported both
in space and time. This means that this solution describes
excitation in some compact space domain (in fact, the ball) $I$,
which exists only some period of time $J$, and vanishes outside.

\section{Appendix: Ultrametric analysis}

In this Section we put the results on ultrametric analysis, which
mainly may be found in \cite{ACHA}, \cite{Izv}. We discuss here
ultrametric wavelet analysis and analysis of ultrametric
pseudodifferential operators (PDO).

\begin{definition}{\sl
An ultrametric space is a metric space with the ultrametric $|xy|$
(where $|xy|$ is called the distance between $x$ and $y$), i.e.
the function of two variables, satisfying the properties of
positivity and non degeneracy
$$
|xy|\ge 0,\qquad |xy|=0\quad \Longrightarrow\quad x=y;
$$
symmetricity
$$
|xy|=|yx|;
$$
and the strong triangle inequality
$$
|xy|\le{\rm max }(|xz|,|yz|),\qquad \forall z.
$$

We say that the ultrametric space $X$ has the analytic type if it
satisfies the following properties:

\medskip

1) The set of all the balls of nonzero diameter in $X$ is no more
than countable;

\medskip

2) For any decreasing sequence of balls $\{D^{(k)}\}$,
$D^{(k)}\supset D^{(k+1)}$, diameters of the balls tend to zero;

\medskip

3) Any ball is a finite union of maximal subballs.

}
\end{definition}

Remind that a directed set is a partially ordered set, where for
any pair of elements there exists the unique supremum with respect
to the partial order.

Denote ${\cal T}(X)$ the set of balls of nonzero diameter in
analytic ultrametric space $X$. Consider the set $X\bigcup {\cal
T}(X)$. This set is directed. The direction if defined by
inclusion of balls and inclusion of points into balls. In
particular, the supremum
$$
{\rm sup}(x,y)=I
$$
of points $x,y\in X$ is the minimal ball $I$, containing the both
points.

Consider a Borel $\sigma$--additive measure $\nu$ with a countable
basis on analytic ultrametric space $X$, such that for arbitrary
ball $D$ its measure $\nu(D)$ is a positive number (i.e. is not
equal to zero).

Consider a basis of of ultrametric wavelets in the space
$L^2(X,\nu)$ of quadratically integrable with respect to the
measure $\nu$ functions. This is a generalization of basis of
$p$--adic wavelets \cite{wavelets}. Generalization of $p$--adic
wavelets onto the family of abelian locally compact groups was
performed by J.J.Benedetto and R.L.Benedetto \cite{Benedetto}.

Denote $V_{I}$ the space of functions on the absolute, generated
by characteristic functions of the maximal subballs in the ball
$I$ of nonzero radius. Correspondingly, $V^0_{I}$ is the subspace
of codimension 1 in $V_{I}$ of functions with zero mean with
respect to the measure $\nu$. Spaces $V^0_{I}$ for the different
$I$ are orthogonal. Dimension of the space $V^0_{I}$ is equal to
$p_I-1$.

We introduce in the space $V^0_{I}$ some orthonormal basis
$\{\psi_{Ij}\}$, $j=1,\dots,p-1$. The next theorem shows how to
construct the orthonormal basis in $L^2(X,\nu)$, taking the union
of bases $\{\psi_{Ij}\}$ in spaces $V^0_{I}$ over all non minimal
$I$.

\begin{theorem}\label{basisX}
{\sl 1) Let the ultrametric space $X$ contains an increasing
sequence of embedded balls with infinitely increasing measure.
Then the set of functions $\{\psi_{Ij}\}$, where $I$ runs over all
non minimal vertices of the tree ${\cal T}$, $j=1,\dots,p_I-1$ is
an orthonormal basis in $L^2(X,\nu)$.

2) Let for the ultrametric space $X$ there exists the supremum of
measures of the balls, which is equal to $A$. Then the set of
functions $\{\psi_{Ij}, A^{-{1\over 2}}\}$, where $I$ runs over
all non minimal vertices of the tree ${\cal T}$, $j=1,\dots,p_I-1$
is an orthonormal basis in $L^2(X,\nu)$.
 }
\end{theorem}

The introduced in the present theorem basis we call the basis of
ultrametric wavelets.

\bigskip

We study the ultrametric pseudodifferential operator (or the PDO)
of the form considered in \cite{ACHA}, \cite{Izv}
$$
Tf(x)=\int T{({\rm sup}(x,y))}(f(x)-f(y))d\nu(y)
$$
Here $T{(I)}$ is some nonnegative function on the tree ${\cal T}(
X)$. Thus the structure of this operator is determined by the
direction on $X\bigcup {\cal T}(X)$. This kind of ultrametric PDO
we call {\it the {\rm sup}--operators} (or operators of the
sup--type).

The next theorem shows that the basis of ultrametric wavelets is
the basis of eigenvectors for ultrametric PDO of the sup--type.

\begin{theorem}\label{04}{\sl Let the following series converge:
\be\label{seriesconverge} \sum_{J>R} T{(J)} (\nu(J)-\nu({J-1,R}))
<\infty \ee where $R$ is some ball in $X$. Then the operator
$$
Tf(x)=\int T{({\rm sup}(x,y))}(f(x)-f(y))d\nu(y)
$$
is self--adjoint, has the dense domain in $L^2(X,\nu)$, and is
diagonal in the basis of ultrametric wavelets from the theorem
\ref{basisX}: \be\label{lemma2.1} T\psi_{Ij}(x)=\lambda_I
\psi_{Ij}(x) \ee with the eigenvalues: \be\label{lemma4}
\lambda_{I}=T{(I)} \nu(I)+\sum_{J>I} T{(J)} (\nu(J)-\nu({J-1,I}))
\ee Here $(J-1,I)$ is the maximal subball in  $J$ which contains
$I$.

Also the operator $T$ kills constants. }
\end{theorem}

\bigskip

\centerline{\bf Acknowledgements}

\medskip

The authors would like to thank I.V.Volovich and B.Dragovich for
fruitful discussions and valuable comments. This paper has been
partly supported by EU-Network ''Quantum Probability and
Applications''. One of the authors (S.K.) has been partly
supported by DFG Project 436 RUS 113/809/0-1, The Russian
Foundation for Basic Research (project 05-01-00884-a), by the
grant of the President of Russian Federation for the support of
scientific schools NSh 1542.2003.1, by the Program of the
Department of Mathematics of Russian Academy of Science ''Modern
problems of theoretical mathematics'', and by the grant of The
Swedish Royal Academy of Sciences on collaboration with scientists
of former Soviet Union.

\end{document}